\documentclass[aps,twocolumn,prd,showpacs,showkeys,preprintnumbers,superscriptaddress,nobibnotes,floatfix,longbibliography,nofootinbib]{revtex4-1}

\pdfoutput=1
\usepackage{amsmath}
\usepackage{amsfonts}
\usepackage{amssymb}
\usepackage{mathrsfs}
\usepackage{graphicx}
\usepackage{xcolor}
\usepackage{longtable}
\usepackage{hyperref}
\usepackage{multirow}
\usepackage{slashed}

\usepackage{array}

\newcommand{\Nuc}[2]{\ensuremath{^{#2}\mbox{#1}}}

\begin{document}

\title{Pushing the frontier of WIMPy inelastic dark matter: 
\\ journey to the end of the periodic table}

\author{Ningqiang Song}
\email{ningqiang.song@queensu.ca}
\affiliation{Department of Physics, Enigneering Physics and Astronomy, Queen's University, Kingston, ON, K7L 3N6, Canada}
\affiliation{Arthur B. McDonald Canadian Astroparticle Physics Research Institute, Department of Physics, Engineering Physics and Astronomy, Queen's University, Kingston ON K7L 3N6, Canada}
\affiliation{Perimeter Institute for Theoretical Physics, Waterloo ON N2L 2Y5, Canada}

\author{Serge Nagorny}
\email{sn65@queensu.ca}
\affiliation{Department of Physics, Enigneering Physics and Astronomy, Queen's University, Kingston, ON, K7L 3N6, Canada}
\affiliation{Arthur B. McDonald Canadian Astroparticle Physics Research Institute, Department of Physics, Engineering Physics and Astronomy, Queen's University, Kingston ON K7L 3N6, Canada}

\author{Aaron C. Vincent}
\email{aaron.vincent@queensu.ca}
\affiliation{Department of Physics, Enigneering Physics and Astronomy, Queen's University, Kingston, ON, K7L 3N6, Canada}
\affiliation{Arthur B. McDonald Canadian Astroparticle Physics Research Institute, Department of Physics, Engineering Physics and Astronomy, Queen's University, Kingston ON K7L 3N6, Canada}
\affiliation{Perimeter Institute for Theoretical Physics, Waterloo ON N2L 2Y5, Canada}

\date{\today}

\begin{abstract}
We explore the reach of low-background experiments made of small quantities of heavy nuclear isotopes in probing the parameter space of inelastic dark matter that is kinematically inaccessible to classic direct detection experiments. Through inelastic scattering with target nuclei, dark matter can yield a signal either via nuclear recoil  or nuclear excitation. We present new results based on this approach, using  data from low-energy gamma quanta searches in low-background experiments with Hf and Os metal samples, and measurements  with CaWO$_4$ and PbWO$_4$ crystals as scintillating bolometers. We place novel bounds on WIMPy inelastic dark matter up to mass splittings of about 640~keV, and provide forecasts for the reach of future experiments.

\end{abstract}

\maketitle

\section{Introduction}
Dark matter (DM) direct detection experiments are designed to record the small amounts of energy expected to be deposited by galactic DM particles as they pass through the detector and interact with nuclei or electrons in the target material. Most such detectors are optimised to search for WIMP-like particles, with masses in the GeV to TeV range, and weak-scale DM-nucleon elastic scattering cross sections $\sigma_n \lesssim 10^{-40}$ cm$^2$. These typically employ target nuclei such as C, O, Si, Ar, Ge or Xe, which are kinematically well-matched to the 10-100 GeV range. Experiments are typically conducted deep underground to mitigate cosmic ray and radiogenic background components.

However, evidence for the existence of DM on large scales does not directly constrain its mass range or interaction strength, but rather its mass density (about 0.3 times the critical density $\rho_c$ on large scales, and around 0.4 GeV cm$^{-3}$ in the Solar neighbourhood), and interaction rate with the Standard Model (SM) of particle physics.
As the viable parameter space for WIMP-like dark matter shrinks following null measurements from sensitive direct detection experiments developed over the past decades, the scope of DM direct detection searches has broadened. Candidates include  sub-GeV dark matter~\cite{Abdelhameed:2019hmk,Liu:2019kzq,Armengaud:2019kfj,Akerib:2018hck,Aprile:2019jmx,Aprile:2019xxb,Agnes:2018oej,Cheng:2021fqb,Barak:2020fql,Aguilar-Arevalo:2019wdi,Arnaud:2020svb,Amaral:2020ryn}, superheavy dark matter~\cite{Bhoonah:2018gjb,Bramante:2019fhi,Garani:2019rcb,Bai:2020ttp,Cappiello:2020lbk,Acevedo:2020gro,Bhoonah:2020fys} and macro dark matter~\cite{Sidhu:2019oii,Sidhu:2019fgg,Sidhu:2019gwo,Bai:2018dxf}. However, the WIMP family still includes many viable candidates that remain out of reach from the current generation of experiments. In particlular, \textit{Inelastic} (I)DM particles, with $O$(keV) or larger mass splittings are  expected in a variety of dark matter models~\cite{Hall:1997ah,TuckerSmith:2001hy,TuckerSmith:2004jv,Arina:2007tm,Cui:2009xq,Fox:2010bu,An:2011uq,Pospelov:2013nea,Dienes:2014via,Barello:2014uda}. Multi-state DM has been invoked in possible explanations of the 511~keV gamma ray excess observed in the galactic center~\cite{Finkbeiner:2007kk,Rea:2007xd,Cline:2010kv,Cline:2012yx,Ema:2020fit} and the DAMA/LIBRA annual modulation signals~\cite{TuckerSmith:2001hy,Chang:2008gd,TuckerSmith:2002af,TuckerSmith:2002af,Kang:2019uuj,Jacobsen:2021vbr}. Recently, interest in inelastic dark matter has also been revived to interpret the XENON1T electron recoil data~\cite{Aprile:2020tmw} with exothermic scattering or luminous dark matter~\cite{Bramante:2020zos,Harigaya:2020ckz,Borah:2020jzi,Choudhury:2020xui,Borah:2020smw,Aboubrahim:2020iwb,Lee:2020wmh,Baek:2020owl,Baryakhtar:2020rwy,Chao:2020yro,An:2020tcg,Lin:2020nxy,Keung:2020uew,He:2020sat,Choi:2020ysq,Dutta:2021wbn}.

A simple inelastic dark matter model consists of a lighter state $\chi_1$ and a heavier state $\chi_2$ with a mass splitting $\delta \equiv M_{\chi_2} - M_{\chi_1}$. 
Generically, the lighter state will dominate the relic abundance, and two-to-two scattering processes with baryons must lead to an excitation or de-excitation of the DM state. In contrast to elastic scattering, the kinematics of inelastic dark matter exhibits two features. First, the dark matter kinetic energy must be large enough to overcome the mass splitting, imposing a lower limit on the relative velocity required for an interaction to take place. Second, the 
minimum momentum exchange required for a collision to occur increases with the mass splitting, meaning that IDM with typical halo velocities can only scatter with heavier nuclei and deposit large enough recoil energy. In other words, even with sufficient kinetic energy, there is a momentum transfer threshold $\sim \mu_{\chi N} v$  below which interactions may not occur. As a consequence, inelastic dark matter may evade the searches in most direct detection experiments, either because the analysis region is limited to low ($\lesssim 10-50$ keV) recoil energies, or because of the limited target nucleus mass.

\begin{figure}
    \centering
    \includegraphics[width=0.9\columnwidth]{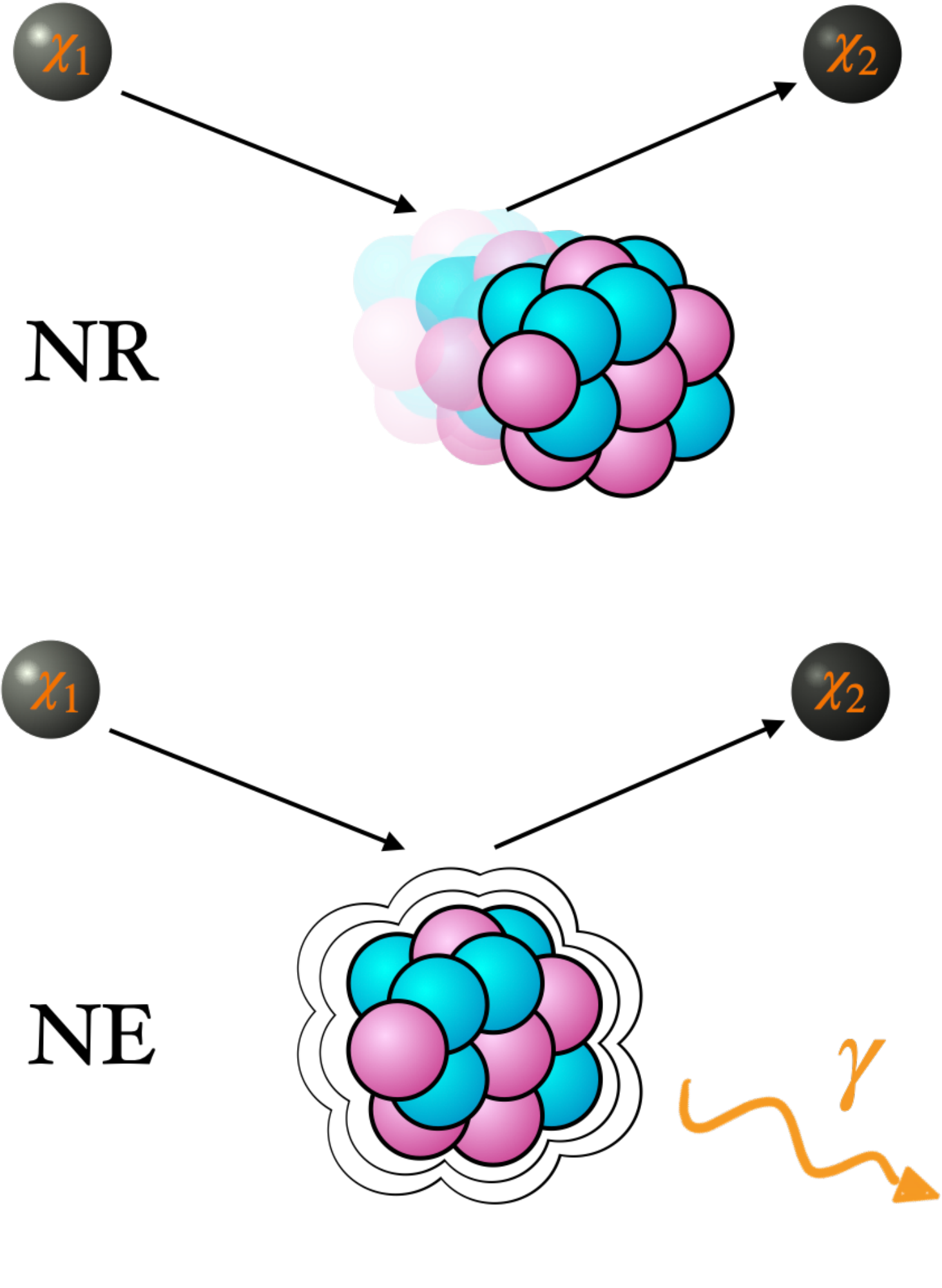}
    \caption{The two processes considered in this work. \textit{Nuclear Recoil} (NR): dark matter is up-scattered to a heavier state and the target nucleus  \textbf{recoils} from the impact, leading to heat, scintillation, and/or ionization signals. \textit{Nuclear Excitation} (NE): in addition to NR, the nucleus can also be \textbf{excited} and then deexcite to ground state with the emission of one or several gamma quanta.}
    \label{fig:DMillustration}
\end{figure}

We thus turn to heavier target nuclei ($Z > 54$), which have received little attention in the search for DM-induced recoils. Many elements between $Z\sim 58$ (cerium) and $Z \sim 83$ (bismuth) are sensitive to IDM models with $\gtrsim$ 100~keV mass splittings, while remaining sufficiently stable not to overwhelm a detector with alpha or beta decay backgrounds. 
Such elements, in particular tungsten and lead, have been utilized in scintillating crystals to search for rare alpha decays~\cite{Munster:2014mga,Beeman:2012wz}. These experiments feature both light and heat measurements, which provide a powerful tool to distinguish DM-induced \textbf{\textit{nuclear recoil}} (NR) events due to high momentum-transfer coherent DM-nucleus interactions (Figure~\ref{fig:DMillustration} top) from background events.

In addition to nuclear recoil signals from coherent scattering, heavier elements can contribute a second type of signal, thanks to their comparatively loosely bound outer nucleon shells: \textbf{\textit{gamma photons}} from induced \textbf{\textit{nuclear excitation }} (NE, Figure~\ref{fig:DMillustration} bottom) and subsequent decay, or deexcitation of a long-lived excited nuclear state. The goal of this paper is thus to use these two approaches, i.e. nuclear recoil searches and nuclear transition gammas to place novel constraints on inelastic DM models.

Recently, a search for collisional deexcitation of the metastable nuclear isomer \Nuc{Ta}{180m} induced by DM has led to novel constraints on  IDM~\cite{Pospelov:2019vuf,lehnert2019search}. \Nuc{Ta}{180m} is fairly unique: it is stable on time scales much larger than the age of the Universe, and  its natural abundance is well established. As the isomers of other nuclei are short-lived, this method is not generally applicable. 
However, the excitation of ground state nuclei to excited states remains possible in the scattering $\chi_1+N\rightarrow \chi_2+N^*$ if the excitation energy is low enough~\cite{Baudis:2013bba,McCabe:2015eia,XENON:2020fgj}. We focus on a few isotopes, with possible nuclear transitions that we will explore listed in  Table~\ref{tab:excitations} below. Once scattered into an excited state, the gamma quanta emitted during subsequent decay to the ground state can be measured. By measuring the activity of these transition lines, we may obtain a conservative upper bound on the total inelastic scattering rate from DM interactions, and thus on the DM-nucleon scattering cross section.  In comparison with traditional direct detection experiments, this method appears rather insensitive. Indeed, nuclear excitation is more kinematically suppressed, and the scattering does not enjoy the usual $A^2$ enhancement factor because the DM is interacting with a single nucleon rather than scattering coherently with the full nucleus. However, we
identify two advantages: 1) The high mass of transition nuclei better matches heavier DM candidates, and 2) the energy of the outgoing gamma is independent of the recoil energy, since it is fixed by the  structure of the target nucleus. This means that, for a given transition, there is no danger of the signal being lost outside of the analysis region of interest. Previously, we have performed such an analysis in the context of a search for rare decays of Hf~\cite{Broerman:2020hfj} and set stronger bounds on IDM mass splittings than previously-reported results from direct detection experiments. In this paper, we extend our approach, using previously-published data collected within low-background measurements with Hf and Os metal samples, along with the limits derived from bolometeric measurements using CaWO$_4$ and PbWO$_4$ scintillating crystals. Although the NE bounds that we find are not as strong as the NR bounds that we derive from bolometer measurements, we will identify a path to leading constraints on WIMPy inelastic dark matter using the nuclear properties of target materials.

This paper is organized as follows. In Sec.~\ref{sec:kinematics} we present the kinematics of inelastic dark matter scattering, including dark matter-induced nuclear transitions and their nuclear response. A number of experiments are listed in Sec.~\ref{sec:experiments} and we show the resulting bounds on inelastic dark matter in Sec.~\ref{sec:results}. Finally, we conclude in Sec.~\ref{sec:conclusions}.
\begin{table*}[!htb]
\centering
\setlength\extrarowheight{3pt}
\begin{tabular}{ c  c  c  c  c  c  c  c}
\hline\hline
Isotope	& Abund.[\%]&$J^p_{\rm g.s.}$  & $J^p_{\rm e.s.}$	& $\Delta E$[keV] & B(E2)[{\it W.u.}] & $\eta$[\%] & Bkg.[mBq/kg]\\ 
\Nuc{Hf}{177} & 18.60 & $7/2^-$ & $9/2^-$ & 112.9500 & 282(8)~\cite{Kondev:2003whm} & 9.64 & 0.9 \\
\Nuc{Hf}{178} & 27.28 & $0^+$ & $2^+$ & 93.1803 & 160(3)~\cite{Achterberg:2009bix} & 7.37 & 2.2 \\
\Nuc{Hf}{180} & 35.08 & $0^+$ & $2^+$ & 93.3240 & 154.8(21)~\cite{McCutchan:2015fnz} & 7.37 & 2.2\\
\cline{3-8}
\multirow{2}{*}{\Nuc{Os}{189}}&\multirow{2}{*}{16.15}&$3/2^-$&$1/2^-$&36.17&27(7)~\cite{Johnson:2017abv} & 0.695&0.40 (proj.)\\
&&$3/2^-$&$5/2^-$&69.54&100(10)~\cite{Johnson:2017abv}&1.75& 0.16\\
\cline{3-8}
\Nuc{Hg}{201}&13.17& $3/2^{-}$ & $1/2^{-}$ & 1.5648 & $\sim 34$~\cite{Kondev:2007suq} & 50 &  0.0056 (proj.)  \\
\hline \hline
\end{tabular}
\caption{Nuclear transitions from ground states (g.s.) to excited states (e.s.) with energy level $\Delta E$ for Hf, Os and Hg isotopes along with their spin, parity and reduced transition probability information. The detection efficiencies $\eta$ near the transition energies are also listed. The most constraining isotopes and transitions are listed with the abundances (see Appendix~\ref{sec:moreisotopes} for a more complete list of transitions and bounds). The gamma background for nuclear excitation (NE) near specific deexcitation energies to the ground states are given at $68\%$ C.L. in the last column. See Section~\ref{sec:NEexperiments} for details.}
\label{tab:excitations}
\end{table*}

\section{Kinematics of inelastic dark matter scattering}
\label{sec:kinematics}
In models of inelastic DM, the dark sector comprises two states $\chi_1$ and $\chi_2$ with a mass splitting $\delta=M_{\chi_2}-M_{\chi_1}$. At tree level, $\chi_2$ couples to $\chi_1$ via a scalar or a gauge boson field, but the self-coupling $\bar{\chi}_1\chi_1$ or $\bar{\chi}_2\chi_2$ is loop-suppressed. In the simplest case, we consider the Lagrangian $\mathcal{L}\supset g\phi \bar{\chi}_2\chi_1+h.c.$. Since $\chi_2$ is heavier than $\chi_1$, after freeze-out we expect $\chi_2$ to be depleted either though decay or through the interconversion process $\chi_2\chi_2\rightarrow \chi_1\chi_1$ when the temperature of the universe falls below the mass splitting, $T<\delta$~\cite{Bramante:2020zos,Batell:2009vb}. As a consequence, the dark matter today is dominated by $\chi_1$. 

We consider two types of dark matter interactions in the detector. The first is nuclear recoil (NR) where the target nucleus displaces, and $\chi_1$ transforms to $\chi_2$. In addition, $\chi_1$ may also interact with the valence nucleons and instigate nuclear excitation (NE) into an unstable state $N^*$. As long as the mediator mass $M_\phi$ is much larger than the momentum transfer  $q$, the interaction can be described by the effective Lagrangian
\begin{equation}
\mathcal{L}\sim \frac{1}{\Lambda^2}\bar{\chi}_2\chi_1\bar{N}^*N,
\end{equation}
where the UV scale $\Lambda$ is set by the mediator mass and the coupling between the dark sector and the SM. We define the quantity $\Delta$, which represents the sum of the DM mass splitting $\delta$ and the excitation energy $\Delta E \equiv E_{N^*}-E_N$:
\begin{equation}
    \Delta\equiv\delta + \Delta E.
\end{equation}
$\Delta E$ vanishes in NR-only scattering.
The kinematics of both interactions follows
\begin{equation}
    \dfrac{\vec{q}_i^2}{2M_{\chi_1}}=\dfrac{\vec{q}_f^2}{2M_{\chi_2}}+\dfrac{\vec{q}^2}{2M_N}+\Delta\,,
    \label{eq:Econservation}
\end{equation}
where the momentum transfer $\vec{q}=\vec{q}_f-\vec{q}_i$. 
Eq.~\eqref{eq:Econservation} can be written
\begin{equation}
    \dfrac{q^2}{2\mu_{\chi N}}-qv\cos\theta+\Delta=0\,,
    \label{eq:qequation}
\end{equation}
where $\mu_{\chi N} \equiv M_{\chi}M_N/(M_{\chi}+M_N)$ is the reduced mass of the DM-nucleus system, and $\theta$ is the laboratory frame scattering angle. We also define $M_\chi\equiv M_{\chi_1}\simeq M_{\chi_2}$ which is valid so long as $M_\chi\gg \delta$. We will assume the dark matter velocity $v$ follows a Maxwellian distribution with $v_0=220$ km/s truncated at the escape velocity $v_{esc}=600$ km/s in the Earth's frame, assuming the Earth's velocity is $v_e=240$ km/s~\cite{monari2018escape}. We refer the reader to Refs.~\cite{Piffl:2013mla,williams2017run,monari2018escape,deason2019local} for detailed discussions of uncertainties on the halo parameter values. The minimum/maximum momentum transfer is obtained at $|\cos\theta|=1$ which gives
\begin{equation}
    q_{\min/\max}=\mu_{\chi N}v\left[1\mp\sqrt{1-2\dfrac{\Delta}{\mu_{\chi N}v^2}}\right]\,,
    \label{eq:qminmax}
\end{equation}
and the minimum/maximum recoil energy is simply obtained via the relationship $E_R=q^2/(2M_N)$. The minimum dark matter velocity required to scatter off a nucleus with {\it any} recoil energy is thus
\begin{equation}
    v_{\min} = \sqrt{2\dfrac{\Delta}{\mu_{\chi N}}}\,.
\end{equation}
Larger dark matter kinetic energy is therefore required for inelastic scattering as $\Delta$ increases. Consequently, the maximum mass splitting that can be probed in an experiment is
\begin{equation}
\delta_{\max}=\dfrac{1}{2}\mu_{\chi N}(v_e+v_{esc})^2-\Delta E\,,
\label{eq:deltamax}
\end{equation}
which grows for larger
reduced mass $\mu_{\chi N}$ and hence heavier target nuclei. For a {\it specific} nuclear recoil energy, the minimum dark matter velocity for an interaction to occur can also be obtained from Eq.~\eqref{eq:qequation}:
\begin{equation}
    v_{\min}(E_R)=\dfrac{1}{\sqrt{2M_N E_R}}\left(\dfrac{M_N}{\mu_{\chi N}}E_R+\Delta\right)\,.
    \label{eq:vminofER}
\end{equation}

\begin{figure}
    \centering
    \hspace*{-.5cm} 
    \includegraphics[width=1.1\columnwidth]{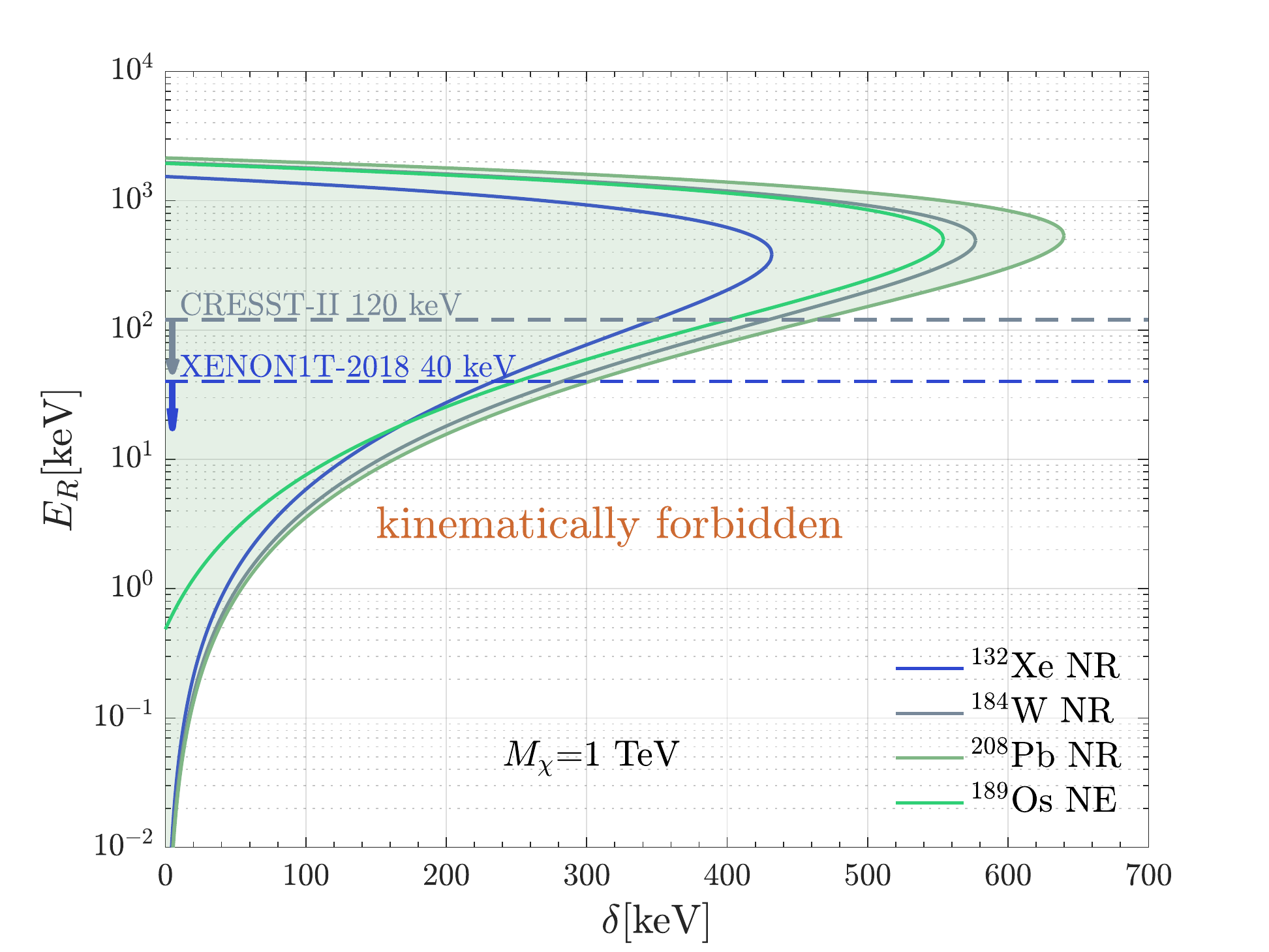}
    \caption{Kinematics of inleastic dark matter with mass splitting $\delta$ scattering with a target nucleus with/without nuclear transition. Kinematically allowed recoil energies are shown for \Nuc{Xe}{132}, \Nuc{W}{184}, \Nuc{Pb}{208} and \Nuc{Os}{189} with the solid lines respectively. The region for \Nuc{Pb}{208} is further shaded. We consider nuclear recoil (NR) for the first three target nuclei, and nuclear excitation (NE) to the 36.17~keV excited state for \Nuc{Os}{189}. The maximum recoil energy analyzed in XENON1T 2018 is about 40~keV and that of CRESST-II~\cite{Angloher:2015ewa} is about 120~keV, as depicted by the horizontal dashed lines. These limit the experiments' respective sensitivities to higher mass splittings. In this figure, the dark matter mass is fixed at $M_{\chi_1}\simeq M_{\chi_2}=1$~TeV.}
    \label{fig:ER}
\end{figure}

The kinematics of NR scattering is shown in Fig.~\ref{fig:ER} for a few nuclei. For each isotope, the region to the left of the corresponding solid curve represents allowed values of $E_R$, given the IDM mass splitting $\delta$. The shape of these curves is determined by the DM velocity distribution. Focusing first on xenon experiments (solid green line), the limitation is twofold: 1) the Xe atomic mass limits the maximum transfer as in  Eq.~\eqref{eq:deltamax}. 2) The analyzed recoil energy, shown here for XENON1T as a dashed blue line, is restricted in most xenon experiments. For the recoil energy range of $1-40$~keV the maximum mass splitting that can be constrained is thus around 250~keV. Recently, XENON1T reported the event spectrum up to the electron recoil energies of 210~keV~\cite{Aprile:2020tmw}. A dedicated analysis of nuclear recoils at high energy would be required to constrain mass splitting above 200~keV, but the sensitivity will still be limited to $\delta<450$~keV due to the limited mass number of Xe isotopes. Conversely, tungsten (solid grey line) is a heavy and stable element, well-suited to exploring high mass splittings. CRESST-II (CaWO$_4$, dashed grey line) analyzed recoil data up to 120~keV, which translates to a mass splitting of about 420~keV. Although the  CRESST-III~\cite{Petricca:2017zdp} data are also available, the detector was optimized to sub-GeV mass dark matter detection with better performance at low recoil energies. Therefore, we still expect CRESST-II to set the leading bound at higher mass splittings.

In addition to NR interactions in traditional direct detection experiments, we are also interested in deexcitation gamma quanta in NE scattering: $\chi_1+N\rightarrow \chi_2+N^*\rightarrow \chi_2+N+\gamma$, whose kinematics is also illustrated in Figure~\ref{fig:ER} for Os. Depending on the IDM model, $\chi_2$ may also decay to SM or invisible particles. If $\chi_2$ decays to photon(s) promptly inside the detector, the decay photon could be resolved as a signature of inelastic scattering; otherwise $\chi_2$ may leave the detector. We conservatively assume the latter. For both NR and NE, the scattering rate is given by
\begin{equation}
    R=\sum\limits_i N_{T_i}\dfrac{\rho_\chi}{M_\chi}\int_{v_{\min}}^{v_{\max}} dv vf(v) \int_{E_{R,{\min}}}^{E_{R,{\max}}} \dfrac{d\sigma_{\chi N}}{dE_R} dE_R\,,
    \label{eq:eventrate}
\end{equation}
where $N_{T_i}$ is the number of the target nuclei of a given isotope in the detector, $\rho_\chi=0.4$ GeV\,cm$^{-3}$ is the local dark matter density, and $f(v)$ is the one-dimensional Maxwellian velocity distribution. The minimum and maximum recoil energies can be found from Eq.~\eqref{eq:qminmax}. The differential cross section is
\begin{equation}
    \dfrac{d\sigma_{\chi N}}{dE_R}=\dfrac{\sigma_n M_N}{2v^2\mu_{\chi n}^2}S(E_R)\,,
\end{equation}
where $\sigma_n$ is the per-nucleon scattering cross section and the nuclear response is encapsulated in $S(E_R)$. For NR, this is just the Helm form factor. In the case of NE, dark matter only interacts with the few valence nucleons on the surface shell of the nucleus, where the nuclear response depends on the momentum transfer without the $A^2$ enhancement factor in $S(E_R)$. As first derived in~\cite{Engel:1992bf}, the nuclear response function is written as
\begin{equation}
    S(\vec{q})=\sum\limits_L|\langle J_f || j_L(qr)Y_{LM}(\hat{r})||J_i\rangle|^2\,,
    \label{eq:Sq}
\end{equation}
where the sum runs over all allowed {\it even} angular momentum states within the range $|J_i-J_f|\leq L \leq |J_i+J_f|$. Here, $J_i$ and $J_f$ are respectively the inital and final angular momenta, $j_L$ are spherical Bessel functions and $Y_{LM}$ are spherical harmonics. The calculation of the response function is usually quite involved for heavy nuclei due to the complexity of the valence structure and the deformation of the nucleus. Fortunately, we may use the reduced transition probability in the nuclear $E2$ transition to approximately eliminate the angular matrix element, which is defined as
\begin{equation}
    B(E2)=\dfrac{1}{2J_i+1}|\langle J_f||er^2Y_2||J_i\rangle|^2\,,
\end{equation}
with $Y_2$ the spherical tensor operator. Assuming the nuclear transition probability density is peaked near the surface of the nucleus, we find~\cite{Engel:1999kv}
\begin{equation}
    S(\vec{q}) = \dfrac{A^2}{Z^2}(2J_i+1)j_2(qR)^2\dfrac{B(E2)}{e^2R^4}\,,
    \label{eq:SE2}
\end{equation}
with the nucleus radius $R\simeq1.2\times A^{1/3}$~fm. The measured reduced transition probabilities $B(E2)$ are listed in Table~\ref{tab:excitations}. These are expressed in terms of the {\it Weisskopf unit} (W.u.)~\cite{Baglin:2008hsa} defined as~\cite{Suhonen:2007zza}
\begin{equation}
    B_{W.u.}(E\lambda)=\dfrac{1.2^{2\lambda}}{4\pi}\left(\dfrac{3}{\lambda+3}\right)^2A^{2\lambda/3}e^2\mathrm{fm}^{2\lambda}\,,
\end{equation}
for a general $E\lambda$ transition. Since higher multipole transitions involving $j_L^2$ are strongly suppressed, we only consider the leading contribution to the nuclear response where $L=2$.

\section{Experiments}

\label{sec:experiments}
Apart from a few notable exceptions, there have been no dedicated experiments searching for DM-nucleus interactions in heavy elements $Z \gtrsim 54$. However, these isotopes have been considered in many calibration experiments as well as searches for rare nuclear decays with the objective of better understanding their nuclear properties. Here, we list a number of promising target compounds, along with published data. We separate these out into two categories: NR, leading to nuclear recoils identifiable in scintillating bolometers, and NE, using gamma ray detectors.

Our strategy is to identify events which could have resulted from a DM-nucleus interaction, and translate the corresponding rate into a limit on the DM interaction rate. This is a conservative approach which, in the absence of a detailed background model for each experiment, can only be used for limit-setting rather than discovery. We will recast these constraints on the rates into constraints on IDM model parameters in Sec. \ref{sec:results}. 

\subsection{Nuclear recoil measurements}
{\it CaWO$_4$.} Crystals used by the CRESST (II-III) \cite{Angloher:2017zkf,Bertoldo:2019xuf} experiments contain a sizable fraction of tungsten, and are thus a good target for IDM searches. However, the analysis region for the CRESST DM runs was limited to low recoil energies, so we  turn to the results of the  crystal characterization runs. The radiopurity of a variety of CaWO$_4$ crystals was investigated at the Gran Sasso National Laboratory (LNGS). We employ measurements between 2009 and 2011 with the CRESST main setup where the scintillation light and phonon signals were both collected. The radiopurity results for a single crystal named \textit{Daisy} are shown in Figure 2 of~\cite{Munster:2014mga}, representing 90.10~kg$\cdot$days of exposure. The two-channel readout (scintillation and phonon) facilitates background event discrimination. The phonon channel was calibrated with $\alpha$ particles, which typically produce a light yield that is around 22$\%$ of the $\beta/\gamma$ light yield in CaWO$_4$ crystals. 
Even less light is expected for nuclear recoils: for example, the light yield for nuclear recoils from a neutron calibration run of CRESST using CaWO$_4$ was found to be below $11\%$~\cite{Angloher:2011uu}. Conservatively, we will set bounds by counting all events observed in \textit{Daisy} below the $\alpha-$band as dark matter candidates. Although the nuclear recoil energy is not available due to the lack of dedicated neutron calibration, we approximate it by the equivalent $\alpha$ energy up to an uncertainty of about 20\%, which does not impact our results in a notable way. With this criterion we identify three dark matter event candidates between 300~keV and 2~MeV, which yield an upper limit of 6.7 events at 90$\%$ CL. Since the bolometric detector is sensitive to the total energy deposition in the heat channel, the efficiency is close to 1 in the energy range we consider.

{\it PbWO$_4$.} Similarly, the $\alpha$ decay of lead isotopes was studied with  11.09 kg$\cdot$days of PbWO$_4$ background measurements, by using the crystal as a scintillating bolometer at the LNGS~\cite{Beeman:2012wz}. A lower light yield is expected from crystals of this type, which for $\alpha$ particles can be estimated via~\cite{Beeman:2012wz}
\begin{equation}
    \frac{LY_{\alpha}}{(\mathrm{keV/MeV})} = (0.28\pm 0.01)+(2.93\pm0.14)\cdot 10^{-5}\cdot \left(\frac{E_\alpha}{\rm keV}\right)\,,
    \label{eq:LY_PbWO4}
\end{equation}
where $E_\alpha$ is the  $\alpha$ particle energy. Taking the events with a lower light yield than expected from $\alpha$ particles as dark matter candidates, we identify 12 events between 600~keV and 2~MeV, which corresponds to an upper limit of 17.8 events at 90\% CL. Likewise we approximate the nuclear recoil energy by the $\alpha$-equivalent energy since most energy is deposited in the heat channel in both cases. The detection efficiency is also close to 1. The results of CaWO$_4$ and PbWO$_4$ experiments are summarized in Table~\ref{tab:DMcandidates}.
\begin{table}[!htb]
\centering
\setlength\extrarowheight{3pt}
\begin{tabular}{ c  c  c  c }
\hline\hline
	& Exposure [kg$\cdot$day] & Phonon $E$ [keV] & DM events\\ 
CaWO$_4$~\cite{Munster:2014mga} & 90.10 & 300-2000 & 3 \\
PbWO$_4$~\cite{Beeman:2012wz} & 11.09 & 600-2000 & 12\\
\hline \hline
\end{tabular}
\caption{Summary of results from CaWO$_4$ and PbWO$_4$ scintillating bolometer experiments, where the experimental exposure, analyzed energy window and the number of dark matter candidates are listed. See Appendix~\ref{sec:eventselection} for details on DM event selection. }
\label{tab:DMcandidates}
\end{table}

\subsection{Nuclear excitation measurements}
\label{sec:NEexperiments}
\begin{figure*}[htb!]
\centering
	\includegraphics[width=0.9\textwidth]{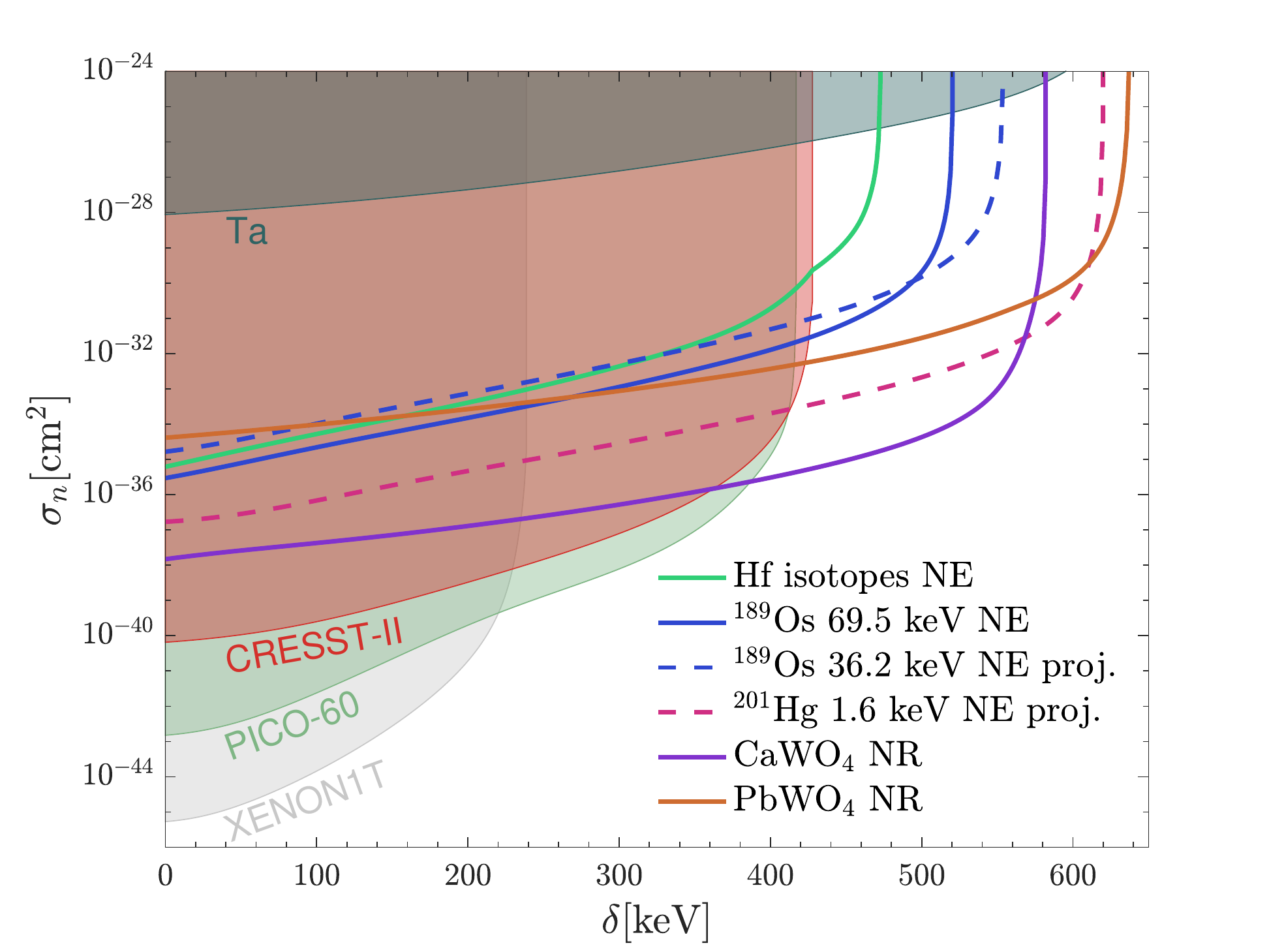}
	\caption{Constraints on inelastic dark matter-nucleon scattering cross section at $90\%$ C.L.. We assume a dark matter mass $M_\chi=1$~TeV, local dark matter density $\rho_\chi=0.4$~GeV/cm$^3$, and a Maxwellian distribution with the Earth velocity $v_e=240$~km/s and the escape velocity $v_{esc}=600$~km/s. Limits based on data from PICO-60~\cite{Amole:2015pla}, CRESST-II~\cite{Angloher:2015ewa}, XENON1T~\cite{Aprile:2018dbl} and \Nuc{Ta}{180} experiment~\cite{lehnert2019search} have been rederived accordingly and are shown with the shaded regions. The green, blue and magenta lines show the cross section limits derived from inelastic dark matter scattering which induces nuclear transitions of Hf~\cite{Broerman:2020hfj}, Os~\cite{Belli:2020vnc}, Hg and the deexcitation gamma rays. Dashed lines are for future projections. The solid purple and orange lines depict the limits on inelastic dark matter from the recoil of heavy nuclei, derived in this work using data from CaWO$_4$~\cite{Munster:2014mga} and PbWO$_4$~\cite{Beeman:2012wz} crystals respectively (see text for details).} 
	\label{fig:dmlimit}
\end{figure*}
{\it Os.} Recently, the search for rare alpha decay of osmium isotopes was conducted at the LNGS~\cite{Belli:2020vnc}. 117.96 grams of high purity Os metal were cut into thin slices to reduce gamma absorption inside the metal. The slices were installed on a Broad-Energy Germanium detector with high sensitivity and ultra-low background. The accumulated energy spectrum was reported, with an exposure of 15,851 hours. The relevant transition levels and the corresponding background rates of Os isotopes are summarized in Table~\ref{tab:excitations}. The deexcitation gamma quanta will appear as a peak in the continuous spectrum. Non-observation of the peak implies that the 68\% C.L. upper limit on the transition rate (marked as ``Bkg'') can be estimated as
\begin{equation}
    R_\mathrm{Bkg}=\sqrt{s\cdot\sigma_\epsilon}/(I_\gamma\cdot\eta\cdot M \cdot t_\mathrm{exp}),
\end{equation}
where $s$ is the accumulated counts per keV as indicated by the measured energy spectrum. The 90\% C.L. limit can be estimated by assuming gaussian distribution, where the transition rate is multiplied by factor of 1.645.  The energy resolution or full width at half maximum (FWHM) is approximated by~\cite{Belli:2020vnc}
\begin{equation}
    \left(\frac{\sigma_\epsilon}{\rm keV}\right)=0.57(5)+0.029(2)\times \sqrt{E_\gamma/\mathrm{keV}}\ \,,
\end{equation}
where $E_\gamma$ is the transition energy. $M$ and $t_\mathrm{exp}$ are given by the mass of an Os isotope and the exposure time respectively. The efficiency of HPGe detector depends on details of the sample including geometry, density and chemical composition, and the energy of gamma quanta. In the absence of the original experimental data, we construct a polynomial fit to the efficiency curve as a function of $\ln(E_\gamma)$~\cite{challan2013gamma} between 46.5~keV and 200~keV, and take the efficiency to be constant below 46.5~keV\footnote{Direct communications with the authors of Ref.~\cite{Belli:2020vnc} show that the 36~keV gamma is below the detection threshold, which we label as ``projection'' in Table~\ref{tab:excitations}.}. A dedicated analysis may yield an $\mathcal{O}(1)$ correction to the efficiency adopted here. If intermediate states exist between the excited nucleus and its ground state, deexcitation may take place through the emission of gamma cascades. Picking a specific gamma energy  for background analysis entails the branching ratio $I_\gamma$, which is 100\% for all transitions listed in Table~\ref{tab:excitations}.  We therefore search for the deexcitation gamma and set bounds on the cross section by requiring that the dark matter scattering rate in Eq.~\eqref{eq:eventrate} not exceed the measured transition rate.

{\it Hf.} Rare decay of hafnium isotopes were also investigated at the LNGS by measuring the internal background of 55.38 grams of Hf foil using modified HPGe detector~\cite{Broerman:2020hfj}. Low-background data has been collected over 310.6 days. The most constraining transitions and the background rates are also listed  in Table~\ref{tab:excitations}. The corresponding limits on inelastic dark matter have been derived in Ref.~\cite{Broerman:2020hfj}.

{\it Hg.} \Nuc{Hg}{201} is an ideal target NE gamma search because of its low-lying excited state  at 1.6~keV. Since Hg is toxic it is preferable to use mercury-containing compounds as the target for dark matter scattering. One such example is to use mercury cadmium telluride (Hg$_{1-x}$Cd$_x$Te).
While this would suffer from a high contamination from \Nuc{Cd}{113} beta decays, which will dominate over other background below 316~keV, increasing the fraction of Hg in the crystal would somewhat  diminish this background component. Conservatively we consider $x=0.2$. The \Nuc{Cd}{113} decay background at 1.6~keV can be obtained by rescaling the experimental data at 100~keV according to the predicted decay spectrum in Fig.~5 of Ref.~\cite{Bodenstein-Dresler:2018dwb}. This 1.6~keV gamma can be measured with an external low-threshold gamma detector such as Ge cryogenic bolometer~\cite{Novati:2019faj,Beeman:2013zva,Barucci:2019ghi}, which is to be installed in close contact with the Hg$_{0.8}$Cd$_{0.2}$Te target. The energy resolution of the Ge gamma detector is estimated to be 50~eV around 1.6~keV~\cite{Barucci:2019ghi}. Assuming 10~kg$\cdot$yr of Hg$_{0.8}$Cd$_{0.2}$Te exposure  and $\sim$ 50\% efficiency for gamma detection, we estimate the background rate to be 0.0004~mBq per kg of  Hg$_{0.8}$Cd$_{0.2}$Te (or equivalently 0.0056~mBq per kg of \Nuc{Hg}{201}) at 1.6~keV transition energy. 

Although the DM analysis in this work is based on the measurement of the 1.6~keV gamma in DM induced NE, the DM limit can be further improved by investigating nuclear recoil signals. Since Hg$_{0.8}$Cd$_{0.2}$Te as a semiconductor is also sensitive to energy deposition, we can improve this setup by measuring the recoil energy of Hg with Hg$_{0.8}$Cd$_{0.2}$Te. The coincident detection of nuclear recoil at Hg$_{0.8}$Cd$_{0.2}$Te and 1.6~keV gamma at Ge bolometer as DM signals provides a powerful way for background rejection. The exploration of this method is left for further work.

\begin{figure*}
    \centering
    \includegraphics[width=0.9\textwidth]{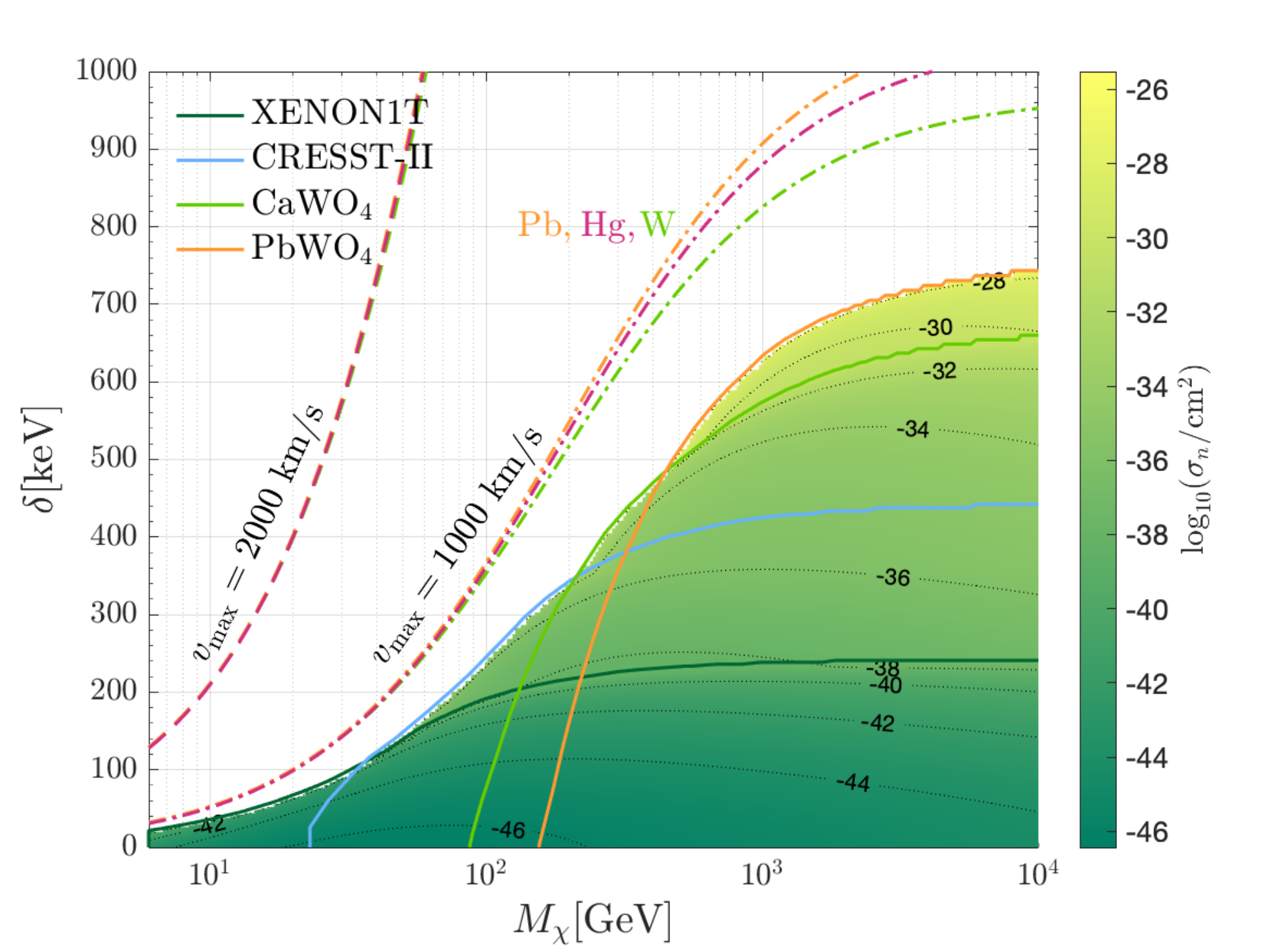}
    \caption{Constraints on inelastic dark matter-nucleon scattering for different dark matter masses and mass splittings. The colored regions to the right of the solid lines are excluded by existing experiments, including XENON1T~\cite{Aprile:2018dbl}, CRESST-II~\cite{Angloher:2015ewa}, and the bolometric searches with CaWO$_4$~\cite{Munster:2014mga} and PbWO$_4$~\cite{Beeman:2012wz} scintillating crystals. Colors and contour lines mark the minimum cross section at $90\%$ C.L. that can be excluded by the combination of experiments. The white region on the upper left is not kinematically accessible in existing experiments. Each experiment is sensitive to a limited amount of parameter space as enclosed by the solid lines. If a subcomponent of the dark matter is boosted to higher velocities, more parameter space can be excluded.  The dash-dotted lines show the projected reach for maximum dark matter velocities of $v_{\max}=1000$~km/s, and the dashed lines for $v_{\max}=2000$~km/s. The green, orange and red broken lines respectively correspond to future dark matter search through scattering with tungsten (nuclear recoil), lead (nuclear recoil) and mercury (nuclear excitation), regardless of detection threshold.}
    \label{fig:inelastic_contour}
\end{figure*}

\section{Results}
\label{sec:results}
The constraints on dark matter-nucleon scattering based on the experiments described above are shown in Figure~\ref{fig:dmlimit}, where we have set the DM mass to $M_\chi=1$~TeV. For dark matter much heavier than the target nuclei, the cross section bound scales linearly with $M_\chi$.

As an illustration of IDM-induced nuclear transition we show the bounds from the NE to different Os excited states in Figure~\ref{fig:dmlimit}. \Nuc{Os}{189} sets the leading bound for mass splittings below 520~keV, due to the collisional excitation to the 9.54~keV state.  Stronger bounds could be obtained from 505~keV to 553~keV by considering the future measurement of 36.17~keV transition gamma . We also set limits on inelastic dark matter by employing Hf rare decay measurements in a similar approach~\cite{Broerman:2020hfj}, which yields less stringent bounds compared with osmium owing to relatively higher transition energies and less massive nuclei. We also propose to detect inelastic dark matter by investigating the deexcitation gamma yield of a mercury-containing compound. With 10~kg$\cdot$yr of HgCdTe exposure as described above, \Nuc{Hg}{201} is expected to yield leading limits amongst nuclear excitation IDM searches for mass splittings between 415~keV and 620~keV.

 We have also derived the bounds for IDM NR scattering on CaWO$_4$ and PbWO$_4$ target crystals acting as scintillating bolometers. Thanks to the simultaneous light and phonon detection, nuclear recoils can be  distinguished on a statistical basis from electron recoil and gamma background. These bounds are more stringent than those from Hf and Os isotopes for all mass splittings. In particular, the CaWO$_4$ experiment dominates the 386~keV to 573~keV range, and the PbWO$_4$ measurements constrain the mass splitting onward up to 636~keV. Nevertheless, \Nuc{Hg}{201} is still expected to lead the 570~keV and 604~keV mass splitting.


We also show the constraints from XENON1T, PICO-60~\cite{Amole:2015pla}, CRESST-II~\cite{Angloher:2015ewa} and \Nuc{Ta}{180m}~\cite{lehnert2019search}. The XENON collaboration reported the results for WIMP searches from about 1 tonne$\cdot$yr of data collection~\cite{Aprile:2018dbl} in 2018. We reproduce the results following the efficiency and events selection in the Xenon1T-2018 code~\cite{xenon2018}. 2 events were observed with 0.9 tonne reference mass and 278.8 days livetime, and the background is expected to be $1.62\pm 0.28$. This translates to an upper limit of 3.7 events at 90\% C.L. Discrimination of nuclear recoil, electron recoil and other backgrounds on an event by event basis requires full likelihood analysis, which is beyond the scope of this work. We derive the bound on inelastic dark matter by integrating over the nuclear recoil energy between 0 and 40.9~keV, with the efficiencies taking care of the low-energy cut.
The PICO-60~\cite{Amole:2015pla} and CRESST-II~\cite{Angloher:2015ewa} bounds in~\cite{Bramante:2016rdh} are recomputed with the dark matter velocity distribution assumed in this paper\footnote{As in~\cite{Bramante:2016rdh} we consider the recoil energy up to 1MeV for PICO and the 30-120~keV recoil energy for CRESST-II.} and the Ta bound~\cite{lehnert2019search} is reproduced with the same assumptions.

We extend the bound to lower and higher dark matter masses in Figure~\ref{fig:inelastic_contour}. The colors depict the smallest cross sections that are excluded with the combination of experiments listed above. The mass splitting that can be constrained rises as dark matter mass increases. The white region is not kinematically accessible due to the limitation of $\mu_{\chi N}$ and dark matter velocity, as is evident from Eq.~\eqref{eq:deltamax}. For $\delta\lesssim 200$~keV leading bounds are set by XENON1T. In the intermediate regime CaWO$_4$ experiment remains the most sensitive search, which is overtaken  by PbWO$_4$ at high mass splittings. Current limits from dark matter induced nuclear excitations remain subdominant. 

Apart from the target nucleus mass, the maximum DM velocity is a limiting factor for probing high mass splittings. Although DM is considered to be virialized, many recent works have shown that a subdominant component could be boosted to high velocity, through cosmic-ray (e.g. \cite{Bringmann:2018cvk,Cappiello:2018hsu,Cappiello:2019qsw,Cho:2020mnc}) or neutrino (e.g. \cite{Yin:2018yjn,Jho:2021rmn,Das:2021lcr}) scattering or stellar acceleration (e.g.~\cite{Emken:2021lgc}), annihilation (e.g.~\cite{Agashe:2014yua,Berger:2014sqa}), semi-annihilation (e.g.~\cite{DEramo:2010keq,Smirnov:2020zwf}) or decay of heavy states (e.g.~\cite{Kong:2014mia,Giudice:2017zke}).  Cosmic-ray upscattered inelastic dark matter with a vector mediator has recently been studied in Ref.~\cite{Bell:2021xff}. Regardless of boost mechanism, we illustrate the kinematic reach of future heavy nuclei experiments to boosted IDM particles in Figure~\ref{fig:inelastic_contour}, for DM velocities of 1000~km/s (dot-dashed lines) and 2000~km/s (dashed lines) assuming W, Pb (NR) and Hg (NE) nuclei as targets. The full parameter space in $M_\chi-\delta$  shown in Fig.~\ref{fig:inelastic_contour} can largely be covered by if $v_{\max}=2000$~km/s. We choose to cap the mass splitting at MeV, beyond which the heavier state may decay to $e^\pm$ in the detector. An exploration of the prompt decay scenario is left for future work.

\section{Conclusions}
\label{sec:conclusions}
As direct detection experiments close in on the window of possible WIMP candidates in the GeV range, inelastic dark matter remains comparatively  unconstrained  by experiments with light nuclei, due to kinematical suppression, as can be seen from Eq.~\eqref{eq:deltamax}. We have extended the scope of this search by investigating nuclear recoil and nuclear excitation events in experiments with heavy nuclei using two distinct experimental approaches.

In the first scenario, target nuclei recoil due to the momentum transferred by dark matter scattering. In a scintillating bolometer,  recoil energies up to about 2~MeV will exhibit distinct features from electron/gamma background due to the low quenching in nuclear recoil. We reinterpret the high phonon energy events with low scintillation yield as dark matter scattering candidate events in experiments with CaWO$_4$ and PbWO$_4$ scintillating bolometers and set bounds on mass splitting up to 636~keV for TeV dark matter. This is the largest splitting explored thus far in the literature. Our limits are conservative as we do not have access to the raw data in the experiments. The discrimination of dark matter candidates from background on an event-by-event basis will strengthen the bounds on inelastic dark matter.  The constraints can be further improved in the future with low-background scintillating crystals~\cite{Pattavina:2019eox}. In particular, RES-NOVA~\cite{RES-NOVA:2021gqp} experiment with large exposure and better background modeling using Monte Carlo, may significantly improve the limits on inelastic dark matter if it is repurposed for dark matter search.  We have also revisited the bounds from XENON1T, PICO-60, CRESST-II and Ta. A likelihood analysis or the study of the high recoil energy data in XENON1T may notably extend the current Xenon bound below 450~keV.

In the second scenario, target nuclei are excited to higher energy states through inelastic collision with dark matter. In the absence of any background process that could yield such an excitation, the deexcitation gamma photons are unique signals of dark matter. In contrast to the difficult measurements with cryogenic scintillating bolometers calibrated at high recoil energies, the detection of nuclear excitation is simply viable with HPGe detector, at the cost of excitation energy and the loss of coherent enhancement in the interaction cross section. We have shown that a recent Os deexcitation gamma search constrains the mass splitting up to 553~keV, and a future Hg experiment could extend the sensitivity up to 620~keV.

A key feature of WIMPy inelastic dark matter resides in its kinematics tightly coupled to the nuclear masses. As we are marching towards the end of the period table, heavy elements are  exploited to tackle the high mass splitting regime, rapidly shrinking this large parameter space. Dedicated high mass, high energy deposition experiments are required to perfect this approach, in combination with more sophisticated analyses of the dark matter flux at different velocities.\\

\section*{acknowledgments}
We thank Joseph Bramante and Peng Wang for helpful discussions, and Harikrishnan Ramani for correspondence on Ta constraints~\cite{lehnert2019search}. We also thank the CRESST group at MPI Munich and Alexander Derbin for openly discussing their Tm data and fruitful collaboration, and Stefano Pirro for communications on Ge detector. SN, NS, and ACV are supported by the Arthur B.~McDonald Canadian Astroparticle Physics Research Institute, with equipment funded by the Canada Foundation for Innovation and the Province of Ontario, and housed at the Queen's Centre for Advanced Computing. Research at Perimeter Institute is supported by the Government of Canada through the Department of Innovation, Science, and Economic Development, and by the Province of Ontario.

\appendix
\section{Bounds from dark matter induced nuclear excitations}

\begin{figure}[htb!]
\centering
	\includegraphics[width=0.48\textwidth]{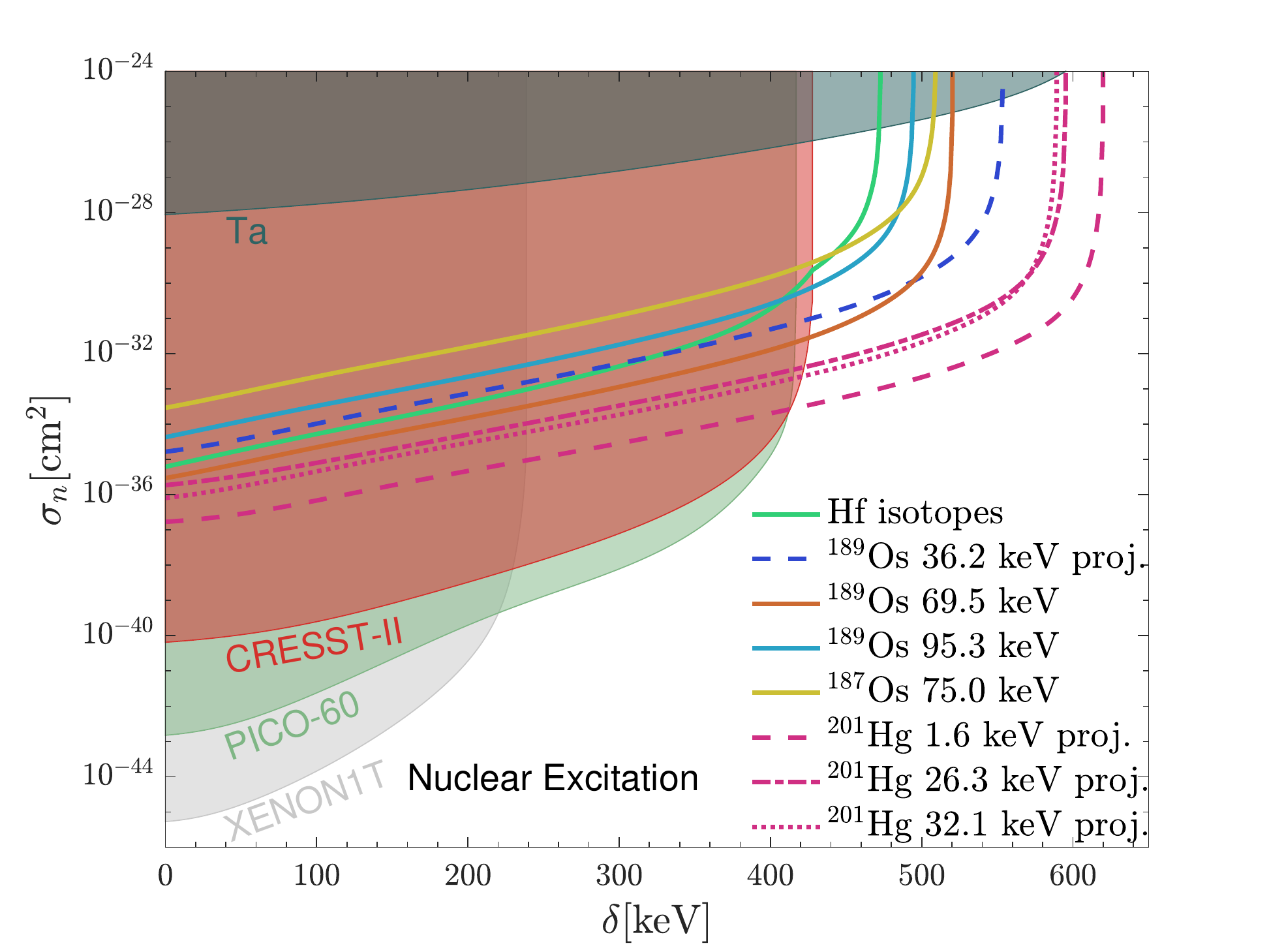}
	\caption{Constraints on inelastic dark matter-nucleon scattering cross section derived from nuclear excitations at $90\%$ C.L.. On top of Figure~\ref{fig:dmlimit} we show in addition the existing bounds from the gamma measurements of the excitation of \Nuc{Os}{189} to 95.3~keV  state and \Nuc{Os}{187} to 75.0~keV state~\cite{Belli:2020vnc}. Projection of nuclear excitation measurement with the \Nuc{Hg}{201} 1.6~keV, 26.3~keV and 32.1~keV excited states and their deexcitations are depicted by dashed, dash-dotted and dotted lines respectively.} 
	\label{fig:dmlimit_allisotopes}
\end{figure}

We explore more isotopes in this appendix. Hf, Os and Hg isotopes and their transitions studied in this work are listed in Table~\ref{tab:excitationsdetail}. Bounds on IDM derived accordingly are shown in Figure~\ref{fig:dmlimit_allisotopes}. For constraints from individual Hf isotope, see~\cite{Broerman:2020hfj}. We do not include \Nuc{Os}{187} 74.4~keV excitation in the analysis since the measured $B(E2)$ is subject to large uncertainty. The projected background rates of transitioning to \Nuc{Hg}{201} 26.3~keV and 32.1~keV excited states are derived similarly to 1.6~keV state by rescaling the \Nuc{Cd}{113} decay data~\cite{Bodenstein-Dresler:2018dwb} and by assuming 50\% efficiency. 

We find the bounds from \Nuc{Os}{187} 75.0~keV and \Nuc{Os}{189} 95.3~keV transitions are inferior to the \Nuc{Os}{189}  69.5~keV constraints. Limits from \Nuc{Hg}{201} 26.3~keV and 32.1~keV are also weaker than 1.6~keV due to the kinematical suppression. Other elements that feature low transition energy and high mass include \Nuc{Ta}{181}, \Nuc{W}{183}, \Nuc{Ir}{193} and \Nuc{Au}{197}, which could be adopted as nuclear targets for future IDM searches.

\label{sec:moreisotopes}
\begin{table*}[!htb]
\centering
\setlength\extrarowheight{3pt}
\begin{tabular}{ c  c  c  c  c  c  c  c c}
\hline\hline
Isotope	& Abund.[\%]&$J^p_{\rm g.s.}$  & $J^p_{\rm e.s.}$	& $\Delta E$[keV] & B(E2)[{\it W.u.}] & $E_\gamma$[keV] & $I_\gamma[\%]$ & Bkg.[mBq/kg]\\
\Nuc{Hf}{174}&0.16&$0^+$&$2^+$&90.985&152(8)~\cite{Browne:1999xcz}&91.00&100&3.8\\ 
\Nuc{Hf}{176}&5.26&$0^+$&$2^+$&88.349&183(7)~\cite{Basunia:2006zql}&88.34&100&3.1\\
\Nuc{Hf}{177} & 18.60 & $7/2^-$ & $9/2^-$ & 112.9500 & 282(8)~\cite{Kondev:2003whm} & 112.9498 & 100 & 0.9 \\
\Nuc{Hf}{178} & 27.28 & $0^+$ & $2^+$ & 93.1803 & 160(3)~\cite{Achterberg:2009bix} & 93.1803 & 100 & 2.2 \\
\Nuc{Hf}{179}&13.62&$9/2^+$&$11/2^+$&122.7904&245(14)~\cite{Baglin:2009mez}&122.793&100&0.9\\
\Nuc{Hf}{180} & 35.08 & $0^+$ & $2^+$ & 93.3240 & 154.8(21)~\cite{McCutchan:2015fnz} & 93.324 & 100 & 2.2\\
\cline{3-9}
\multirow{2}{*}{\Nuc{Os}{187}}&  \multirow{2}{*}{1.8794}&$1/2^-$&$3/2^-$&74.356&$50^{+60}_{-50}$~\cite{Basunia:2009hgg}&74.30 &100& 0.25\\
&&$1/2^-$&$5/2^-$&75.016 & 38(10)~\cite{Basunia:2009hgg}&64.31&54.3& 0.31\\
\cline{3-9}
\multirow{3}{*}{\Nuc{Os}{189}}&\multirow{3}{*}{16.152}&$3/2^-$&$1/2^-$&36.17&27(7)~\cite{Johnson:2017abv} &36.17& 100 & 0.40\\
&&$3/2^-$&$5/2^-$&69.54&100(10)~\cite{Johnson:2017abv}&69.53&99.8& 0.16\\
&&$3/2^-$&$3/2^-$&95.27&14(3)~\cite{Johnson:2017abv} &59.06& 75.5 & 0.20\\ 
\cline{3-9}
\multirow{3}{*}{\Nuc{Hg}{201}}&\multirow{3}{*}{13.17}& $3/2^{-}$ & $1/2^{-}$ & 1.5648 & $\sim 34$~\cite{Kondev:2007suq} &1.5648& 100 & 0.0056 (proj.)  \\
& & $3/2^{-}$ & $5/2^{-}$ & 26.272 & 2.4(8)~\cite{Kondev:2007suq} &26.34& 100 &  0.0060 (proj.)\\
& & $3/2^-$&$3/2^-$&32.145&20(9)~\cite{Kondev:2007suq} &32.19&50.5 & 0.012 (proj.) \\
\hline \hline
\end{tabular}
\caption{Nuclear transitions from ground states (g.s.) to excited states (e.s.) with energy level $\Delta E$ for Hf, Os and Hg isotopes along with their spin, parity, reduced transition probability information and the isotope abundances. The deexcitation gamma energies $E_\gamma$ employed in the analysis are listed with their branching ratio $I_\gamma$ (normalized to 1). The gamma background for nuclear excitation (NE) near specific deexcitation energies to the ground states are given at $68\%$ C.L. in the last column. See Section~\ref{sec:NEexperiments} for details.}
\label{tab:excitationsdetail}
\end{table*}

\section{Event selection for scintillating bolometer experiments}
\label{sec:eventselection}

\begin{figure}[htb!]
\centering
	\includegraphics[width=0.48\textwidth]{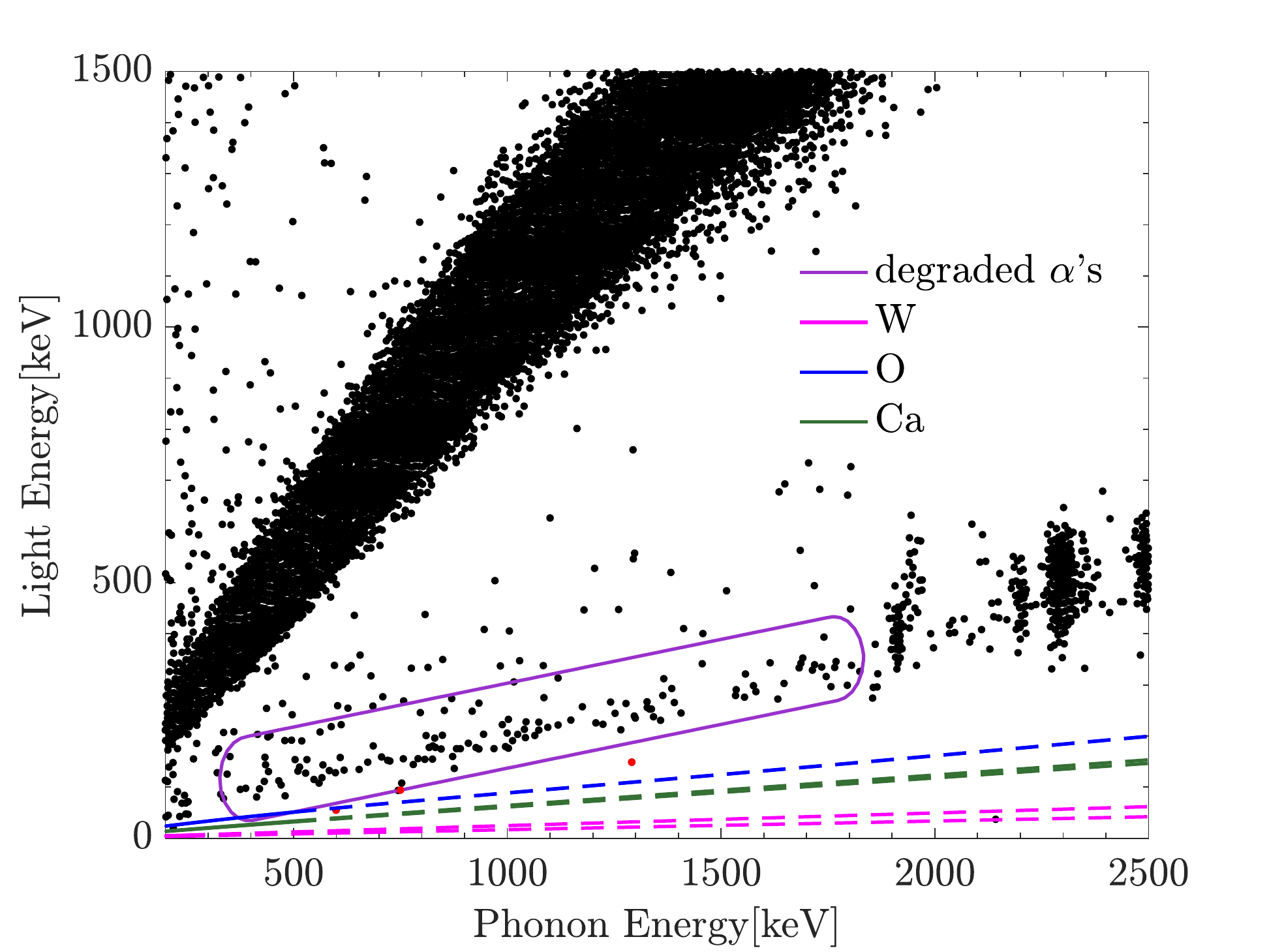}
	\caption{Event discrimination in the CaWO$_4$ scintillating bolometer experiment. The purple band corresponds to the degraded $\alpha$-events in Ref.~\cite{Munster:2014mga}. The blue, and green and magenta lines show the measured light yields of O, Ca and W in nuclear recoil at 90\% C.L.~\cite{strauss2014energy}. The light yield of W recoil is measured up to $\sim$240~keV recoil energy, and that of O and Ca recoils are measured up to $\sim$500~keV, beyond which we depict light yields as dashed lines. Events painted red below the purple alpha-band are considered as dark matter candidates.} 
	\label{fig:DM_CaWO4}
\end{figure}

In Fig.~\ref{fig:DM_CaWO4} we show the dark matter candidate events in the CaWO$_4$ scintillating bolometer experiment. The light yields (LYs) of O, Ca and W in nuclear recoil have been studied in Ref.~\cite{strauss2014energy}, which all approach a constant at high energies. The LYs of O and Ca can be parameterized as
\begin{equation}
    LY(E_R)=LY^\infty (1+fe^{-E_R/\lambda})\,,
\end{equation}
with $LY^\infty$ being the LY at $E_R=\infty$. The relevant parameters are given in Table~\ref{tab:LY}. The LY of W in nuclear recoil is measured to be $0.0208\pm0.0024$. The LYs are depicted as solid and dashed lines in Fig.~\ref{fig:DM_CaWO4} at 90\% C.L. (again gaussian distribution is assumed to infer the 90\% C.L. from $1\sigma$ range), which lie below the $\alpha$-band from Ref.~\cite{Munster:2014mga}. This reconfirms the low LY of nuclear recoil compared with $\alpha$'s. Conservatively we assume the events below the purple $\alpha$-band as dark matter candidates, where three such events are found for the phonon energy between $300$~keV and $2000$~keV. We do not include the events between 1800~keV and 2000~keV, as they are attributable to \Nuc{Nd}{144} decays and are far away from the W LY, which dominates the dark matter scattering at high recoil energy due to kinematical suppression.

\begin{table}[!htb]
\centering
\setlength\extrarowheight{3pt}
\begin{tabular}{ c  c  c  c }
\hline\hline
	& $LY^\infty$ & $f$  & $\lambda$[keV]\\ 
O & $0.07908\pm 0.00002$ & $0.7088 \pm 0.0008$ & $567.1 \pm 0.9$ \\
Ca & $0.05949 \pm 0.00078$ & $0.1887 \pm 0.0022$ & $801.3 \pm 18.8$\\
\hline \hline
\end{tabular}
\caption{$1\sigma$ range of parameters for the light yields of O and Ca taken from Ref.~\cite{strauss2014energy}.}.
\label{tab:LY}
\end{table}

\begin{figure}[htb!]
\centering
	\includegraphics[width=0.48\textwidth]{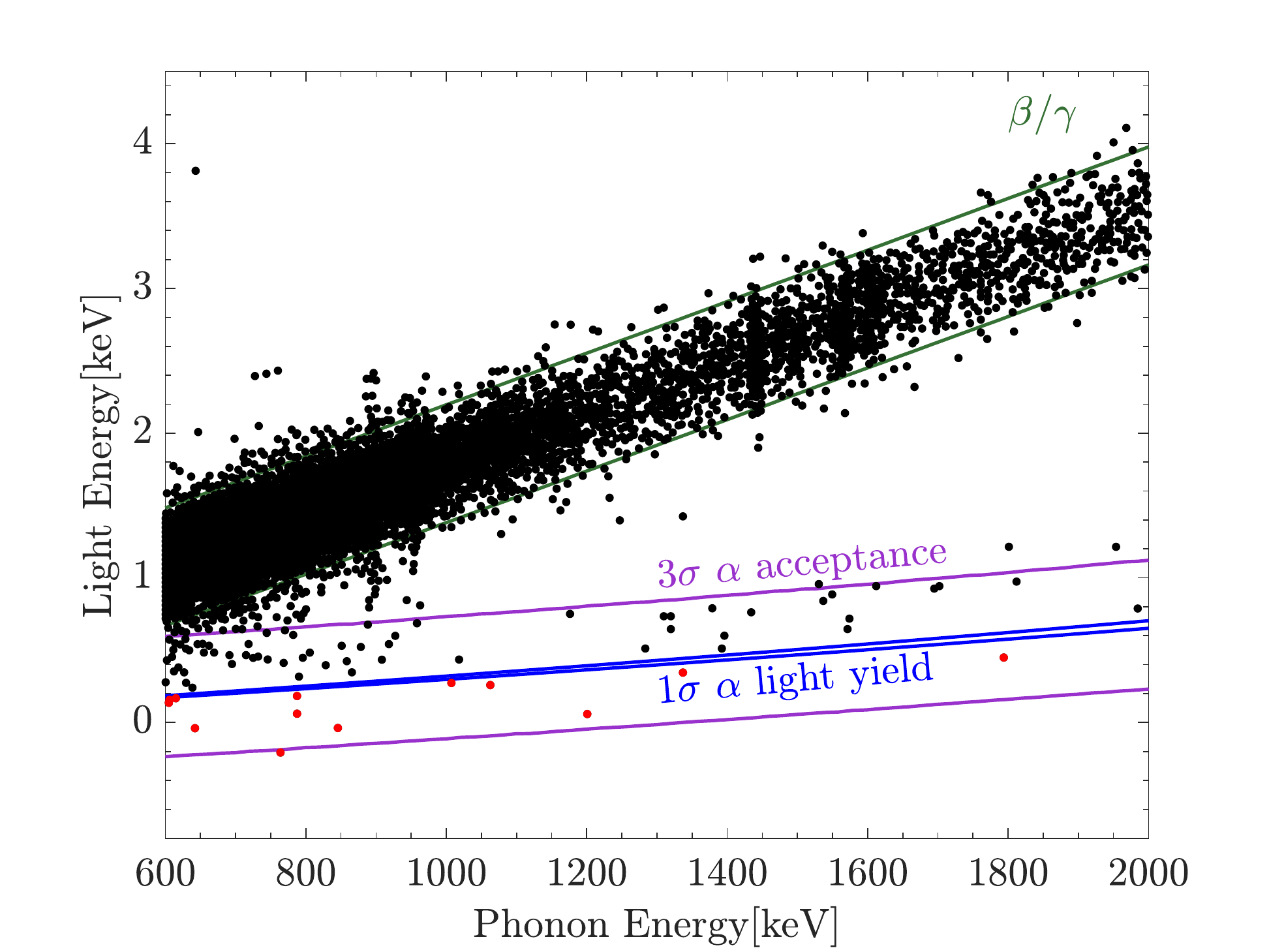}
	\caption{Event discrimination in the PbWO$_4$ scintillating bolometer experiment. The green and purple lines correspond to the 3~$\sigma$ acceptance region for $\beta/\gamma$ and $\alpha$ events adapted from Ref.~\cite{Beeman:2012wz} respectively. The blue lines represent the 1~$\sigma$ region of $\alpha$ events based on the light yield of $\alpha$ particles. Events painted red below the blue lines are considered as dark matter candidates.} 
	\label{fig:DM_PbWO4}
\end{figure}

We also show the event selection in the PbWO$_4$ experiment~\cite{Beeman:2012wz} in Fig.~\ref{fig:DM_PbWO4}. The LY of $\alpha$'s is given in Eq.~\eqref{eq:LY_PbWO4}. We show the $1\sigma$ range of the corresponding LY with blue lines. To be conservative, we assume events below the $1\sigma$ $\alpha$-band to be dark matter candidates, where 12 such events are identified. 

\bibliography{inelastic}

\end{document}